\documentclass[journal=nalefd,manuscript=letter]{achemso}

\usepackage{graphicx}
\usepackage{epstopdf}

\usepackage[hidelinks]{hyperref}
\usepackage{color}
\usepackage{amsmath,amssymb,amstext,graphicx,bm}

\author{Lada~Vuku\v{s}i\'c}
\email{lada.vukusic@ist.ac.at}
\phone{+43 2243 9000 721204}
\affiliation[IST Austria]{Institute of Science and Technology Austria, Am Campus 1, 3400 Klosterneuburg, Austria}

\author{Josip~Kuku\v{c}ka}
\affiliation[IST Austria]{Institute of Science and Technology Austria, Am Campus 1, 3400 Klosterneuburg, Austria}

\author{Hannes~Watzinger}
\affiliation[IST Austria]{Institute of Science and Technology Austria, Am Campus 1, 3400 Klosterneuburg, Austria}

\author{Joshua Michael Milem}
\affiliation[IST Austria]{Institute of Science and Technology Austria, Am Campus 1, 3400 Klosterneuburg, Austria}

\author{Friedrich Sch\"{a}ffler}
\affiliation{Johannes Kepler University, Institute of Semiconductor and Solid State Physics, Altenbergerstr 69, 4040 Linz, Austria}

\author{Georgios Katsaros}
\affiliation[IST Austria]{Institute of Science and Technology Austria, Am Campus 1, 3400 Klosterneuburg, Austria}

\title{Single-shot readout of hole spins in Ge}

\begin{document}


\begin{abstract}
The strong atomistic spin orbit coupling of holes makes single-shot spin readout measurements difficult because it reduces the spin lifetimes. By integrating the charge sensor into a high bandwidth radio-frequency reflectometry setup we were able to demonstrate single-shot readout of a germanium quantum dot hole spin and measure the spin lifetime. Hole spin relaxation times of about 90 $\mu$s at 500\,mT are reported. By analysing separately the spin-to-charge conversion and charge readout fidelities insight into the processes limiting the visibilities of hole spins has been obtained. The analyses suggest that high hole visibilities are feasible at realistic experimental conditions underlying the  potential of hole spins for the realization of viable qubit devices.
\end{abstract}

\newpage

\section{Introduction}

Spin-based qubit systems have been in the focus of intense research in the past 15 years \cite{Hanson2007,ZwanenburgReview2013}, showing continuous improvement in the coherence times \cite{Muhonen2014} and quality factor, the ratio between the qubit coherence and manipulation time \cite{Yoneda18}. One of the requirements for the realization of any type of qubit is a readout mechanism with high fidelity \cite{DiVincenzoCriteria}. For spin 1/2 qubits and in single quantum dot devices this is realized optically by means of luminescence measurements \cite{WarburtonReview} and electrically by spin to charge conversion. The later was introduced in 2004 for electrons in GaAs \cite{Elzerman2004}. A few years later, a similar scheme was used in order to measure the spin relaxation times for electrons in Si \cite{MorelloNature2010,Buech13}. However, so far there has been no demonstration of single-shot hole spin readout despite the fact that holes are becoming more and more attractive as viable qubits \cite{Maurand2016,PrechtelNatMat2016,Watzinger18} and have shown promising spin relaxation times \cite{Hu2012,GerardotNature2008,HeissPRB2007}.

Here we study hole quantum dots (QDs) formed in Ge hut wires (HWs) \cite{Zhang2012} and we demonstrate for the first time single-shot hole spin readout. Due to the strong spin orbit coupling \cite{HaoNanoLett2010, HigginbothamPRL2014,Kloeffel2017}, which in general leads to shorter relaxation times \cite{Hanson2007}, we integrated the charge sensor into a radio-frequency reflectometry setup \cite{Schoelkopf1998}. Such a setup allows high bandwidths and the extraction of hole spin relaxation times which were measured to be about 90 $\mu$s at 500mT.

HWs are an appealing platform for building quantum devices with rich physics and technological potential. The confined hole wavefunction is almost of purely heavy-hole character  \cite{Watzinger2016} which can lead to long spin coherence times \cite{Fisher2008}. Furthermore, they are monolithically grown on Si \cite{Zhang2012} without the use of any catalyst making them fully compatible with CMOS technology. In addition, as self-assembled nanostructures can be grown on prepatterned Si substrates \cite{KatsarosPRL2008,ZhangAPL2007} one can envision the growth of HWs at predefined positions.  



\begin{figure}[hhhhhhh!!!]
\includegraphics{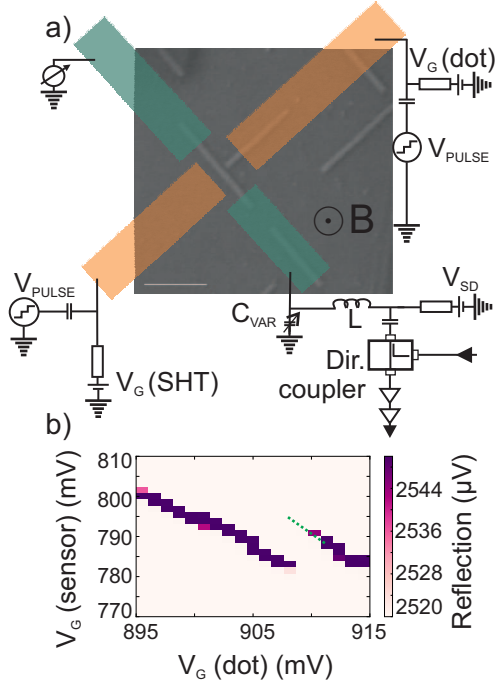}%
\caption{\label{fig:figure1} Spin readout device and schematics. (a) Schematic of a device similar to one used for the spin readout with the scanning electron micrograph of the HWs in the background. Source and drain electrodes are shown in green, gates in orange. The scale bar  is 200\,nm. (b) Zoom-in of a stability diagram obtained by sweeping the gate of the QD versus the gate of the charge sensor, at a magnetic field of 1100\,mT. The pulsing sequence was applied along the upper part of the Coulomb peak break (green dashed line).}
\end{figure}

\begin{figure*}[tbp]
\includegraphics{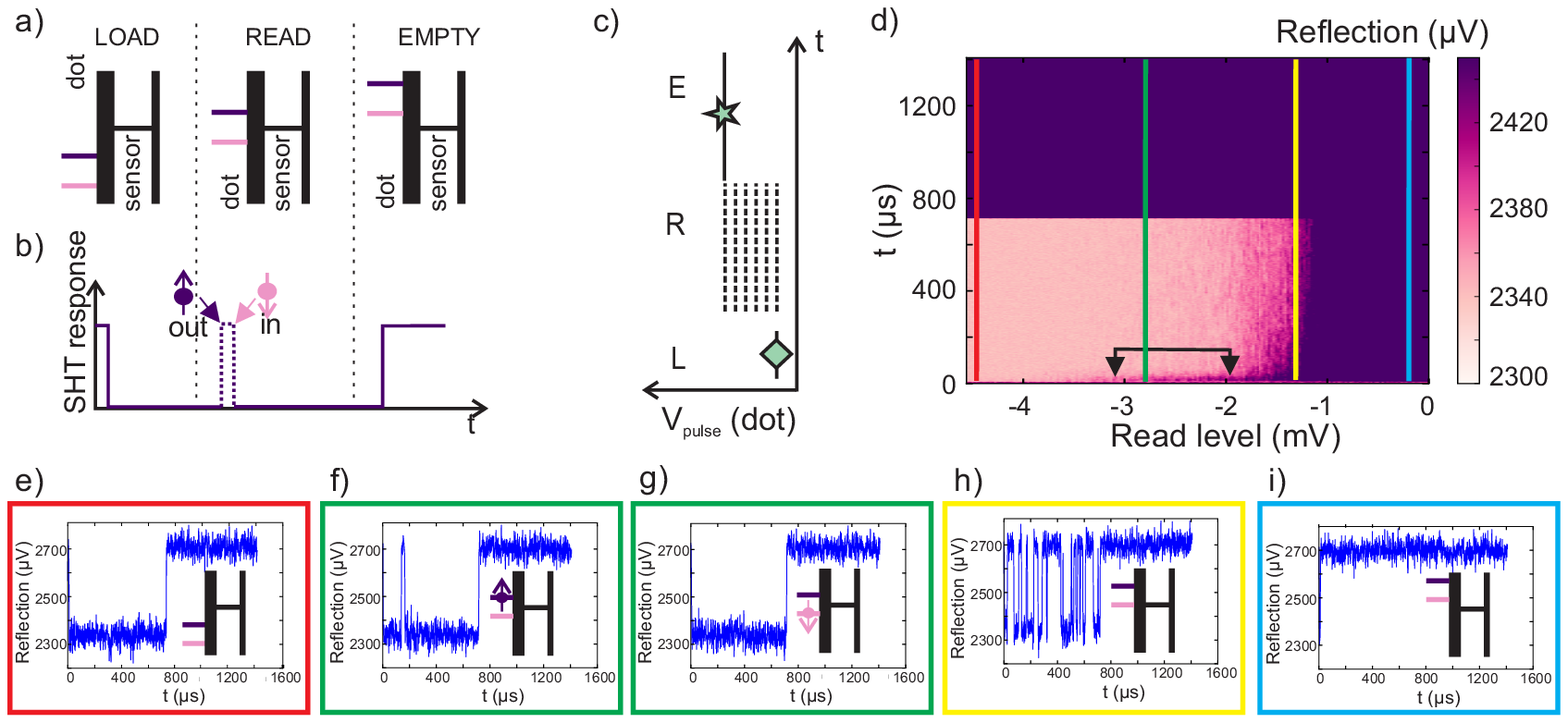}%
\caption{\label{fig:figure2} Single-shot spin readout and calibration of the read level.  (a) Schematics showing the electrochemical potentials of the QD and the charge sensor during different stages of the pulsing sequence used for the single-shot spin readout. The lower electrochemical potential corresponds to a spin down state. For simplicity throughout the manuscript the electron convention is used in the diagrams showing the alignment of the electrochemical potentials. (b) Expected response of the SHT when the sequence is applied along the upper part of the Coulomb peak break and a spin up hole is loaded. (c) Three-stage pulsing sequence. The duration of the load stage is 8\,$\mu$s and that of the read and empty stages 700\,$\mu$s. (d) Reflection amplitude, averaged over 197 single-shot traces as a function of the voltage applied on the QD gate during the read stage, taken at the magnetic field B\,=\,1100\,mT, with a detection bandwidth of 200\,kHz. The double black arrow indicates the region where we see the spin signature.  (e)-(i) Examples of single-shot traces; the schematics in the insets elucidate the alignment of the electrochemical potentials at the positions indicated by vertical lines in (d). (e) The read level is set too low: $\mu_{\uparrow}, \mu_{\downarrow} < \mu_{SHT}$, no hole can leave the QD during the read stage. (f) Correct position of the read level: $\mu_{\downarrow} < \mu_{SHT} < \mu_{\uparrow}$. Single-shot trace for the case of loading a spin up hole. (g) Correct position of the read level: $\mu_{\downarrow} < \mu_{SHT} < \mu_{\uparrow}$. Single-shot trace for the case of loading a spin down hole. (h)  $\mu_{\downarrow} \approx \mu_{SHT}$. Random telegraph signal showing the continuous exchange of holes between the QD and the SHT.  (i) The read level is set too high: $\mu_{\uparrow}, \mu_{\downarrow} > \mu_{SHT}$: the hole can always tunnel out during the read stage.}
\end{figure*}

The device used in this study consists of a QD formed at the end of a Ge HW and a charge sensor capacitively and tunnel coupled to it, which is used both as a hole reservoir and for the spin readout \cite{VukusicNano}. The charge sensor is a single hole transistor (SHT), formed in a HW which grows perpendicular to that hosting the spin qubit (Fig. \ref{fig:figure1}a). Whenever a hole tunnels from the QD to the charge sensor a break in the SHT coulomb peak appears (Fig. \ref{fig:figure1}b). In the presence of an external magnetic field, such a single hole tunnelling event becomes spin selective. In order to detect it, the Zeeman splitting, $E_Z = g\mu_BB$ must be larger than the width of the Fermi distribution of the SHT states; where g denotes the g-factor, $\mu_B$ the Bohr magneton and B the applied magnetic field.

For performing single-shot measurements with high bandwidth, we used a reflectometry-based readout setup, where the SHT is part of the resonant circuit \cite{ReillyAPL07,PettaAPL12, BuitelaarGroup2015, AresPRA16,HileAPL2015}. A radio frequency (RF) wave is sent towards the SHT and each change in its impedance manifests as a change in the amplitude of the reflected wave. All measurements were performed in a dilution refrigerator with a base temperature of $\approx$15\,mK.

\section{Results}

For the spin readout measurement we use the already well established three-stage pulsing sequence (Fig. \ref{fig:figure2}a) implemented by Elzerman et al. \cite{Elzerman2004} to do spin-to-charge conversion. In a first stage (\textit{load}), a hole with an unknown spin is loaded from the sensor into the dot. In a second stage (\textit{read}), the electrochemical potentials of the QD for spin up ($\mu_{\uparrow}$) and spin down ($\mu_{\downarrow}$) are brought in a configuration where $\mu_{\uparrow}$  is above and $\mu_{\downarrow}$ below the electrochemical potential of the SHT ($\mu_{SHT}$). With the last pulse (\textit{empty}), the loaded hole tunnels out of the QD. The charge sensor, SHT, shows maximum (minimum) reflection amplitude (RA) when the QD is empty (loaded) (Fig. \ref{fig:figure2}b). In the read phase one distinguishes between two cases, depending on whether a spin up or spin down hole has been loaded. In case a spin down hole is loaded, the SHT RA stays at its minimum during the read stage. However, when a spin up hole is loaded it can tunnel out of the QD. As a consequence the SHT RA obtains its maximum value until it switches back to the minimum value when the QD gets refilled with a spin down hole.

For determining, in the first place, the correct position of the read level for which spin dependent tunnelling is occurring, a similar three-stage sequence was applied (Fig. \ref{fig:figure2}c), with the difference that the amplitude of the read stage was varied. Averaging about 200 single-shot measurements reveals the spin signature (Fig. \ref{fig:figure2}d) as a purple tail at the beginning of the read phase between roughly -3\,mV and -2\,mV (black double arrow in Fig. \ref{fig:figure2}d). Different RA responses of the SHT are observed depending on the position of the read level, starting from too low (Fig. \ref{fig:figure2}e) to too high (Fig. \ref{fig:figure2}i). The green line in Fig. \ref{fig:figure2}d is positioned such that $\mu_{\downarrow} < \mu_{SHT} < \mu_{\uparrow}$. Two single-shot measurements taken at the position of the green line are shown in Fig. \ref{fig:figure2}f) and Fig. \ref{fig:figure2}g). Fig. \ref{fig:figure2}f corresponds to a loaded spin up hole, while Fig. \ref{fig:figure2}g to a spin down hole. For the neighbouring break of the same Coulomb peak we do not see the spin signature, as this method works only when the QD has an even number of holes before the load stage. We note that in our measurements we could not see the existence of discrete energy levels in the SHT.
\begin{figure}[tbp]
\includegraphics[scale=0.95]{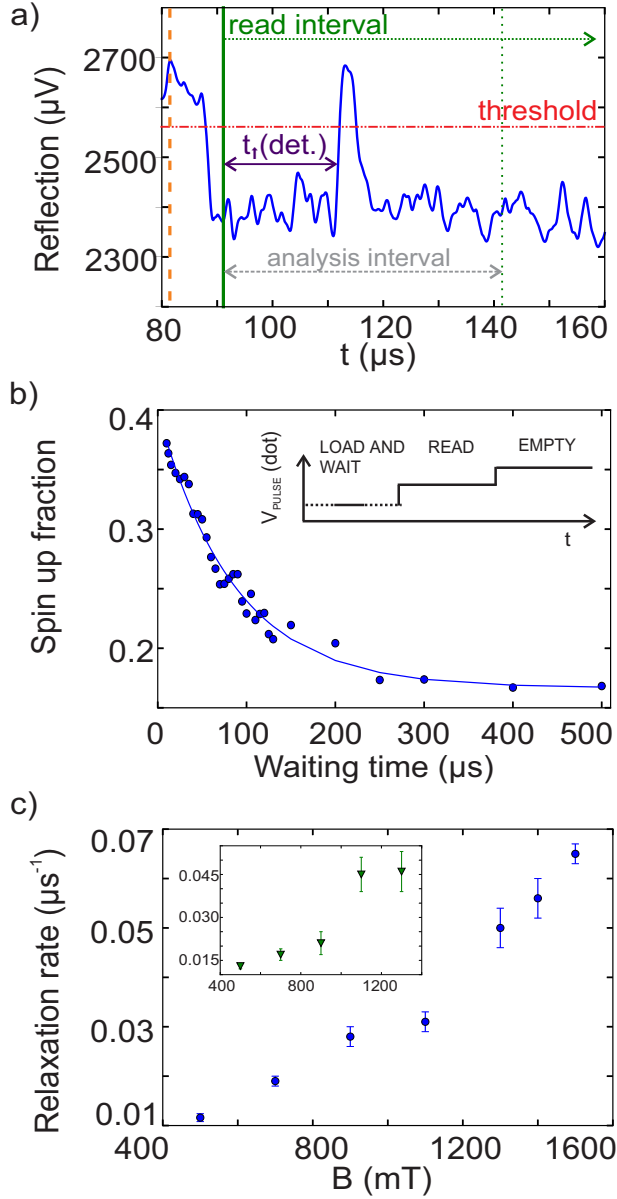}%
\caption{\label{fig:figure3} Spin relaxation rate. (a) Example of a single-shot trace for a loading time of 10\,$\mu$s and for a magnetic field of 500\,mT. The beginning of the load stage is labelled with the vertical dashed orange line and the moment when the levels of the dot are pulsed to the read stage with the vertical solid green line. The horizontal dot-dashed red line indicates the threshold above which a tunnelling event is considered to have taken place. All single-shot analysis was performed for an interval of 50$\mu$s (grey dashed double arrow), as after the 50$\mu$s and for tunnelling times of about 10$\mu$s, the number of counts for spin up tunnelling-out events is less than 1\%. (b) Exponential decay of the spin up fraction versus the waiting time for  B\,=\,500\,mT. The three stage pulsing sequence for measuring the spin relaxation time is shown in the inset. The duration of both the read and the empty stage is 700\,$\mu$s and the duration of the load stage was varied from 10\,$\mu$s to 500\,$\mu$s. (c) Plot showing the spin relaxation rate vs magnetic field. The results for a second measured Coulomb peak break are shown in the inset. For the first break the QD confines about 10\,-\,20 holes while the second break corresponds to approximately 10\, holes less.}
\end{figure}

Once the correct position of the read level was determined, the sequence for spin readout was applied (Fig. \ref{fig:figure3}b, inset). In order to extract the hole spin relaxation time, the duration of the first, load stage of the pulse, is varied, while the durations of the read and empty stages are kept constant. The probability of observing a spin-up hole decreases exponentially with the waiting time. From the exponential decay, we extract a hole spin relaxation time $T_1$ of $(86 \pm 6)$\,$\mu$s for out-of-plane magnetic fields of 500\,mT (Fig. \ref{fig:figure3}b). As expected, the spin relaxation rate $T_1^{-1}$ increases when increasing the magnetic field $B$ (Fig. \ref{fig:figure3}c). We note that the values extracted from the single-shot measurements are in agreement with those extracted by integrating the averaged RA \cite{Supplementary}.

\begin{figure}[tbp]
\includegraphics[scale=0.95]{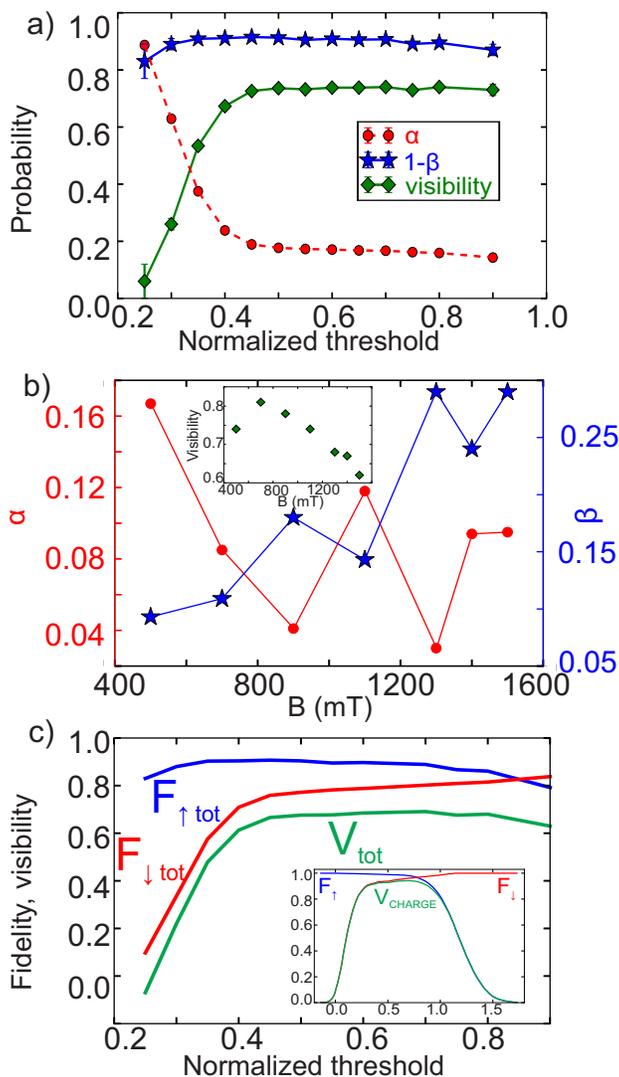}%
\caption{\label{fig:figure4} Measurement fidelity. (a) Dependence of $\alpha$, 1-$\beta$ and the visibility on the normalized threshold, at B\,=\,500\,mT. Threshold\,=\,1 corresponds to the average maximum SHT RA and threshold\,=\,0 to the average minimum SHT RA. (b) Magnetic field dependence of $\alpha$ (red dots) and $\beta$ (blue stars), extracted for the threshold which gives the maximum visibility. The inset shows the dependence of the spin-to-charge visibility vs B. (c) Plot showing the total readout spin down ($F_{\downarrow tot}$\,=\,(1-$\alpha$)*$F_{\downarrow}$, red), spin up ($F_{\uparrow tot}$\,=\,(1-$\beta$)*$F_{\uparrow}$, blue) fidelity and the total readout visibility ($F_{\downarrow tot} + F_{\uparrow tot} -1$, green). The inset shows the spin down ($F_{\downarrow}$, red) and spin up ($F_{\uparrow}$, blue) charge readout fidelities, as well as the charge readout visibility (green). The normalized threshold can exceed 1.0 as there are RA values exceeding the average maximum value.} 
\end{figure}

\section{Discussion}

The magnetic field dependence of  $T_1^{-1}$ does not follow a $B^{5}$ curve (Fig. \ref{fig:figure3}c) which has been shown for electrons in GaAs and Si \cite{Hanson2007,MorelloNature2010,Buech13}. For holes the Bir-Pikus Hamiltonian which describes the strain in the valence band needs to be considered \cite{BirPikusBook}. As it contains strain tensor elements and the spin operators \cite{Kloeffel2017} it leads to a hole-phonon Hamiltonian which depends both on the spin and the phonons. This is different from the conduction band where the electron-phonon interaction does not depend on the spin. By taking the Bir-Pikus Hamiltonian into account a $B^{7/2}$ hole spin relaxation rate dependence has been recently predicted for Ge/Si core/shell nanowires \cite{Maier2013PRB}. However, the experimental data reveal a $B^{1.5}$ for the first break and a $B^{1.4}$ dependence of the spin relaxation rate \cite{Supplementary} which is deviating from what was predicted from theory for cylindrical nanowires.

To estimate the accuracy of the single-shot spin readout measurements, we followed a hybrid approach based on the methods introduced by Elzerman et al. \cite{Elzerman2004} and Morello et al. \cite{MorelloNature2010}. This approach allowed us to get insight in the limitations of hole spins as potential qubits. Initially we extracted the spin-to-charge conversion fidelities. For each threshold used in the single-shot analysis, we extracted two parameters, $\alpha$ and $\beta$. Both correspond to a wrong assignment of the spin states. The parameter $\alpha$ gives the probability that the SHT signal exceeds the threshold even in the case of loading a spin down hole, and can be extracted from the saturation value of the spin up fraction for very long waiting times (Fig. \ref{fig:figure3}b). The parameter $\beta$ corresponds to the probability that a spin up hole relaxes before it tunnels out. It is equal to 1/(1 + $T_1\Gamma_{\uparrow}$), where $\Gamma_{\uparrow}$ is the spin up tunnel rate. From the fit to the histogram representing the detection times of the spin up hole (t$_\uparrow$(det) in Fig. \ref{fig:figure3}a), one can extract the decay rate equal to $(\Gamma_{\uparrow} + T_1^{-1})$, which then allows the extraction of $\Gamma_{\uparrow}$ \cite{Supplementary}. Due to the large setup bandwidth, $\beta$ is largely threshold insensitive as shown in Fig. \ref{fig:figure4}a) for 500\,mT. The spin-to-charge conversion fidelity for the spin down hole (1-$\alpha$) is 0.833\,$\pm$\,0.005 while for the spin up hole (1-$\beta$) it is 0.907\,$\pm$\,0.007, giving a maximum spin-to-charge conversion visibility (1- $\alpha$ - $\beta$) of 0.740$\pm$\,0.009 for the normalized threshold of 0.7 (Fig. \ref{fig:figure4}a).

In order to get a better understanding into the factors limiting the spin-to-charge conversion fidelities for holes, the dependence of $\alpha$ and $\beta$ on the magnetic field was investigated (Fig. \ref{fig:figure4}b). While $\alpha$ tends to decrease for larger magnetic fields, $\beta$ shows the opposite behaviour. This leads to a maximum total spin-to-charge conversion visibility of 0.81$\pm$\,0.01 at 700\,mT. $\alpha$ implies mainly a failure of the spin down hole to remain in the QD. The tunnel out time of the spin down state and thus 1-$\alpha$ depends on the ratio of the magnetic field and the effective electron temperature (EET) \cite{Buech13}. One solution for increasing 1-$\alpha$ is to increase the magnetic field. However, larger magnetic fields imply short spin relaxation times and large qubit operation frequencies. The optimal solution is to keep the magnetic field at low values and decrease the effective electron temperature. Since the reported experiment was performed at an EET of about 300-400\,mK \cite{Supplementary}, fidelities 1-$\alpha$ higher than 0.95 should be feasible at magnetic fields of about 200\,mT for an EET of 100\,mK. We now turn our attention to $\beta$. As $\Gamma_{\uparrow}$ is rather insensitive to the magnetic field \cite{Supplementary}, the increase of $\beta$ originates from the drastically reduced spin relaxation times. Taking into account the $B^{3/2}$ dependence of the spin relaxation rate, relaxation times exceeding 0.3\,ms should be feasible at 200\,mT. This is in line with the values reported for core-shell wires at low magnetic fields \cite{Hu2012}. Such longer spin relaxation times will allow 1-$\beta$ to exceed 0.95. From the above discussion it becomes clear that the main difference of hole spins compared to electron spins lays in $\beta$. While $\alpha$ for electron spins is as well limited by the magnetic field value \cite{Buech13}, this is not the case for $\beta$. For electron spins the spin relaxation time is in the order of seconds even at fields exceeding 1\,T \cite{MorelloNature2010,Buech13}, which in combination with the short tunnelling times makes $\beta$ rather insensitive to the value of the magnetic field.

We now move to the charge readout fidelity. For this we performed a simulation following the procedure introduced by Morello et al. \cite{MorelloNature2010,Supplementary}. Spin down and spin up fidelities of 0.962 and 0.980 were obtained (Fig. \ref{fig:figure4}c, inset). These fidelities are as high as those reported for electron spins as for the charge readout fidelities it is the measurement bandwidth which determines the extracted values. Finally, in order to obtain the total spin up and spin down fidelities and visibility of the single-shot measurements, the spin-to-charge conversion and charge readout fidelities were multiplied (Fig. \ref{fig:figure4}c). The total spin down (up) hole fidelity is given by 0.801\,$\pm$\,0.005 (0.889\,$\pm$\,0.007) and the total visibility of the single shot readout measurements is 0.691\,$\pm$\,0.008. These values correspond to the normalized threshold of 0.7. By repeating the same analysis for 700\,mT a total visibility of 0.752\,$\pm$\,0.009 was obtained \cite{Supplementary}.

In summary, as the interest in hole spin qubits \cite{Maurand2016,Watzinger18} has been continuously increasing over the past few years \cite{Hendrickx2018,Brauns2016PRB,LiNanoLett2015,Wang2016NanoLett}, the demonstration of hole spin readout in single QD devices is an important first step towards more complex geometries \cite{Zajac2018Science,Veldhorst2015Nature,Watson2018Nature}. The reported spin-to-charge conversion and charge readout out fidelities suggest that hole devices operated at low magnetic fields can lead to qubits with spin readout fidelities exceeding 0.9. The reported results together with the CMOS compatibility, the possibility of isotopical purification and the strong spin orbit coupling, suggest Ge as a promising material system for moving towards long range coupling and spin entanglement \cite{Nigg2017PRL, Kloeffel2013PRB}.


\section{Methods}

The Ge HWs used in this study were grown by a solid-source molecular beam epitaxy (MBE) system on 4-inch intrinsic Si(001) wafers via the Stranski-Krastanow growth mechanism. The wafers were dipped in a hydrofluoric solution before loading into the MBE chamber. After degassing at 720\,$^{\circ}$C, a Si buffer layer was deposited. Then, 6.6\,\AA \,of Ge were deposited on the substrate at 580\,$^{\circ}$C  followed by an in-situ annealing of 5h at 570\,$^{\circ}$C . The amount of the deposited Ge is at the critical thickness for the nucleation of three dimensional hut clusters. At last, the temperature was decreased to 300\,$^{\circ}$C and the samples were capped with 4\,nm of Si.

Electrical contacts on the wires were fabricated by means of electron-beam lithography and electron-beam metal evaporation. For the source and the drain electrodes a combination of Al/Pd was used (7\,nm/27\,nm).  Before the metal deposition, a short dip in buffered hydrofluoric acid was performed in order to remove the native oxide. For the gate electrodes Ti/Pd (5/20\,nm) were deposited on top of  $\approx$8\,nm hafnium oxide created by atomic layer deposition.

All measurements were performed in a dilution refrigerator with a base temperature of 15 mK. A Tektronix
AWG5014C arbitrary wave generator was used to apply voltage pulses to the gates and the Zurich Instruments
UHFLI lock-in amplifier was used for the readout. The sample was mounted onto a printed circuit board incorporating RC filters (R\,=\,10\,k$\Omega$, C\,=\,10\,nF) for the DC lines, the bias tees for the reflectometry (R\,=\,10\,k$\Omega$, C\,=\,10\,nF) and the fast gate lines (R\,=\,1.8\,M$\Omega$, C\,=\,10\,nF). The matching circuit consisted of an 820\,nH inductor (1206CS-821XJLB) and a varactor (MA46H070-1056) which was biased with 3\,V. The fast gate and the input reflectometry lines were attenuated by 27\,dB and 42\,dB, respectively. Attenuators were distributed at the different stages of the dilution refrigerator
and at room temperature. The reflected signal was amplified at two stages, once at 4K and once at room
temperature. We used a CITLF2 cryogenic amplifier, a ZX60-33LN-S+ room temperature amplifier and a ZX30-9-4-S+ directional coupler. The power of the RF signal on the lock-in output was -35\,dBm.

\section{Data availability}

The data that support the findings of this study are available from the corresponding author upon reasonable request.

\section{Acknowledgments}
We thank C. Kloeffel, A. Laucht, D. Loss and M. Veldhorst for helpful discussions.
The work was supported by the ERC Starting Grant no. 335497, the FWF-Y 715-N30
project and the Austrian Ministry of Science through the HRSM call 2016. This research was supported by the Scientic Service Units of IST Austria through resources provided by the MIBA Machine Shop and the Nanofabrication Facility. 

\section{Contributions}

L.V. performed the measurements and analysed the data under the supervision of G.K.
J.K. contributed to the development of the RF parts for the experiment. L.V. fabricated
the devices with participation of H.W. and G.K.. H.W. and F.S. were responsible for the growth of the samples. J.M.M. performed the simulation for the charge readout fidelity. L.V. and G.K. designed the experiment and wrote the manuscript with input from all authors

\section{Competing Interests}
Authors declare no competing interests.


\end{document}